# Continuation-Passing Style, Defunctionalization, Accumulations, and Associativity

Jeremy Gibbons[a]

a Department of Computer Science, University of Oxford, UK

**Abstract**

**Context** Reynolds showed us how to use continuation-passing style and defunctionalization to transform a recursive interpreter for a language into an abstract machine for programs in that language. The same techniques explain other programming tricks, including zippers and accumulating parameters.

**Inquiry** Buried within all those applications there is usually a hidden appeal to the algebraic property of associativity.

**Approach** Our purpose in this paper is to entice associativity out of the shadows and into the limelight.

**Knowledge** We revisit some well-known applications (factorial, fast reverse, tree flattening, and a compiler for a simple expression language) to spotlight their dependence on associativity.

**Grounding** We replay developments of these programs through a series of program transformations and data refinements, justified by equational reasoning.

**Importance** Understanding the crucial role played by associativity clarifies when continuation-passing style and defunctionalization can help and when they cannot, and may prompt other applications of these techniques.

# 1 Introduction

In a seminal paper [28], Reynolds showed how to use *continuation-passing style* (CPS) followed by *defunctionalization* to take a direct-style recursive interpreter written in some (presumably well understood) *defining language* implementing some (typically less well understood) *defined language*, and transform it into an abstract machine for executing programs in the defined language. CPS ensures that the order of execution of the resulting machine is independent of that of the defining language, for example whether the defining language is lazy or eager; and defunctionalization ensures that the abstract machine is first-order even if the defining language was higher-order.

Danvy and colleagues have subsequently elaborated on Reynolds' approach in a long sequence of papers [13, 2, 1, 3, 4, 12], using it to reconstruct many published abstract machines and to construct some new ones. One can also see CPS and defunctionalization applied in contexts other than language interpreters and abstract machines—for example, Huet's Zipper [16] and McBride's Clowns and Jokers [22]. Moreover, they shed light on some other standard techniques, such as accumulating parameters [9], translation between right-to-left and left-to-right folds [8, §3.5], and fast reverse [17].

This much is fairly well known. However, behind all these applications of the two techniques is an appeal to associativity, which is not so well known. Perhaps it should not come as a surprise that CPS transformation—which after all is a matter of sequentializing code that would otherwise be tree-structured—is connected to associativity. Nevertheless, we feel that the associativity aspect deserves to be highlighted. The purpose of this paper is to explain and explore the connection between CPS, defunctionalization, and associativity.

The paper uses the dependently typed language Idris as a programming notation, for which a brief primer is given in Appendix A. The paper is mostly independent of any specifics of the language; in particular, the only real reliance on Idris's dependent types is for implementing a generalized form of composition in section 5, and is not essential to the central argument. The source of the paper is a literate script, and the extracted code is provided as supplementary material [14].

# 2 Warm-up: Factorial

We start with a very simple example, in order to set out the main ideas in a familiar context—a straightforward direct-style recursive implementation of factorial:

$$\begin{aligned}
&\mathit{fact} : \mathit{Nat} \to \mathit{Nat} \\
&\mathit{fact}\ Z \quad\ \ = 1 \\
&\mathit{fact}\ (S\ n) = S\ n \times \mathit{fact}\ n
\end{aligned}$$

We deliberately work through the details here, so that we can present the later examples more briskly.

### 2.1 Continuation-passing style

The first step is to convert the direct-style definition to *continuation-passing style*. This is achieved by introducing an additional argument $k$, an *accumulating parameter*, thereby *generalizing* the function *fact* to another function $fact_2'$ specified by

$$
\begin{array}{l}
fact_2' : Nat \to (Nat \to r) \to r \\
fact_2'\ n\ k = k\ (fact\ n) \qquad \text{-- specification of } fact_2'
\end{array}
$$

This really is a generalization, because *fact* may be retrieved by setting $k = id$. Indeed, we define

$$
\begin{array}{l}
fact_2 : Nat \to Nat \\
fact_2\ n = fact_2'\ n\ id
\end{array}
$$

and so $fact_2 = fact$. (As a convention throughout the paper, we use the same name but different indices for different versions of morally the same function. Sometimes the different versions really are extensionally equal; more generally, they may be related by data refinement.)

From the original definition of *fact*, and the specification of $fact_2'$ in terms of it, using simple unfold–fold transformations [10], it is now straightforward to synthesize a definition of $fact_2'$ that no longer depends on *fact*. For the base case, we have

$$
\begin{array}{ll}
 & fact_2'\ Z\ k \\
= & [\![\ \text{specification of } fact_2'\ ]\!] \\
 & k\ (fact\ Z) \\
= & [\![\ \text{definition of } fact\ ]\!] \\
 & k\ 1
\end{array}
$$

and for the inductive step:

$$
\begin{array}{ll}
 & fact_2'\ (S\ n)\ k \\
= & [\![\ \text{specification of } fact_2'\ ]\!] \\
 & k\ (fact\ (S\ n)) \\
= & [\![\ \text{definition of } fact\ ]\!] \\
 & k\ (S\ n \times fact\ n) \\
= & [\![\ \text{composition}\ ]\!] \\
 & (k \cdot (\lambda m \Rightarrow S\ n \times m))\ (fact\ n) \\
= & [\![\ \text{specification of } fact_2'\ ]\!] \\
 & fact_2'\ n\ (k \cdot (\lambda m \Rightarrow S\ n \times m))
\end{array}
$$

We therefore define:

$$
\begin{array}{l}
fact_2' : Nat \to (Nat \to r) \to r \\
fact_2'\ Z\ k \qquad = k\ 1 \\
fact_2'\ (S\ n)\ k = fact_2'\ n\ (k \cdot (\lambda m \Rightarrow S\ n \times m))
\end{array}
$$

This is the continuation-passing style implementation. Clearly the continuation argument $k$ is being used as an accumulating parameter: it starts off as *id*, accumulates more information (here, by function composition) as the computation proceeds, and yields the final result by some projection (here, by application to 1). An equivalent way of writing the second clause of $fact'_2$ is

$$fact'_2\ (S\ n)\ k = fact'_2\ n\ (\lambda m \Rightarrow k\ (S\ n \times m))$$

in which the right-hand side might be read "compute the factorial of $n$; get a result, let's call it $m$; then return $k\ (S\ n \times m)$". This version is more idiomatically in continuation-passing style, and we use this style without comment for later examples.

### 2.2 Defunctionalization

Now $fact'_2$ is *tail-recursive*: when it recurses, the result of the recursive call is also the final result. But it is still higher-order: it manipulates continuations. However, in the context of its use as an auxilliary function to $fact_2$, we don't need the full generality of $fact'_2$: the continuations are always of type $Nat \to Nat$, so we could have given $fact'_2$ the more specific type

$$fact'_2 : Nat \to (Nat \to Nat) \to Nat$$

Moreover, the continuations aren't even arbitrary functions of type $Nat \to Nat$: they're specifically compositions of the form $(a\times)\cdot(b\times)\cdots(c\times)$. Having made this observation, we can choose a different representation of functions in this special form, and transform $fact'_2$ by *data refinement* to use this different representation instead. If we pick a first-order representation, we get a first-order implementation of $fact'_2$. In particular, we can represent a function of the form $(a\times)\cdot(b\times)\cdots(c\times)$—that is, the composition of a possibly empty sequence of multiplications by various factors $a, b, \ldots, c$—by the list $[a, b, \ldots, c]$ of those factors. The *abstraction function* $factabs_3$ converts from the new 'concrete' representation back to the old 'abstract' one:

$$\begin{aligned} &factabs_3 : List\ Nat \to (Nat \to Nat) \\ &factabs_3\ [\,] \qquad\;\; = id \\ &factabs_3\ (n :: k) = (n\times) \cdot factabs_3\ k \end{aligned}$$

—that is, $factabs_3\ k\ m = foldr\ (\times)\ m\ k$. Data refinement then starts from the specification

$$fact'_3\ n\ k = fact'_2\ n\ (factabs_3\ k) \qquad \text{-- specification of } fact'_3$$

and leads to the new implementation

$$\begin{aligned} &fact_3 : Nat \to Nat \\ &fact_3\ n = fact'_3\ n\ [\,]\ \ \textbf{where} \\ &\quad fact'_3 : Nat \to List\ Nat \to Nat \\ &\quad fact'_3\ Z\ k \qquad = foldr\ (\times)\ 1\ k \qquad \text{-- } = product\ k \\ &\quad fact'_3\ (S\ n)\ k = fact'_3\ n\ (k +\!\!+ [S\ n]) \end{aligned}$$

This is still tail-recursive, but is now also first-order.

### 2.3 Associativity

Of course, $fact_3$ is not the tail-recursive program for factorial that you would sit down and write from first principles. That program arises from a further data refinement, representing the composition $(a\times)\cdot(b\times)\cdots(c\times)$ by the product $a\times b\times\cdots\times c$ of factors—a single integer, rather than a list of integers. Now the abstraction function back to the original continuations used by $fact'_2$ is

$$\begin{array}{l} factabs_4 : Nat \to (Nat \to Nat) \\ factabs_4\ k = (k\times) \end{array}$$

Applying the data refinement starts from the specification

$$fact'_4\ n\ k = fact'_2\ n\ (factabs_4\ k) \qquad \text{-- specification of } fact'_4$$

and leads to the new implementation

$$\begin{array}{l} fact_4 : Nat \to Nat \\ fact_4\ n = fact'_4\ n\ 1 \ \ \textbf{where} \\ \quad fact'_4 : Nat \to Nat \to Nat \\ \quad fact'_4\ Z\ k = k \\ \quad fact'_4\ (S\ n)\ k = fact'_4\ n\ (k \times S\ n) \end{array}$$

This is still tail-recursive and first-order, but now also manipulates only scalar data. But note that this final data refinement is *valid only because of associativity of multiplication*. The crucial part of the development is the inductive step in the auxilliary function:

$$\begin{array}{ll} & fact'_4\ (S\ n)\ k \\ = & [\![\ \text{specification of } fact'_4\ ]\!] \\ & fact'_2\ (S\ n)\ (factabs_4\ k) \\ = & [\![\ \text{definition of } fact'_2\ ]\!] \\ & fact'_2\ n\ (\lambda m \Rightarrow factabs_4\ k\ (S\ n \times m)) \\ = & [\![\ \text{definition of } factabs_4\ ]\!] \\ & fact'_2\ n\ (\lambda m \Rightarrow k \times (S\ n \times m)) \\ = & [\![\ \text{associativity of multiplication}\ ]\!] \\ & fact'_2\ n\ (\lambda m \Rightarrow (k \times S\ n) \times m) \\ = & [\![\ \text{definition of } factabs_4\ ]\!] \\ & fact'_2\ n\ (\lambda m \Rightarrow factabs_4\ (k \times S\ n)\ m) \\ = & [\![\ \text{specification of } fact'_4\ ]\!] \\ & fact'_4\ n\ (k \times S\ n) \end{array}$$

Suppose we had started instead with a non-associative operator, as in the 'subtractorial' function:

$$\begin{array}{l} subt : Integer \to Integer \\ subt\ 0 = 1 \\ subt\ n = n - subt\ (n-1) \end{array}$$

We could have followed the same steps to get a tail-recursive $subt_2$ and a first-order $subt_3$, but we would need to do something different in order to get a scalar $subt_4$. It's a nice exercise to work out just what that difference would be—but because subtraction is not associative, a different insight is needed than for factorial.

# 3 Reverse

Let's look at another familiar linear example: list reversal. This goes through similar steps, but the dependence on associativity arises in a different way. We start with the naive, quadratic-time, direct-style reverse function:

$$\begin{aligned}
&\mathit{reverse} : \mathit{List}\ a \to \mathit{List}\ a \\
&\mathit{reverse}\ [\,] \qquad\quad = [\,] \\
&\mathit{reverse}\ (x :: \mathit{xs}) = \mathit{reverse}\ \mathit{xs} +\!\!+ [x]
\end{aligned}$$

## 3.1 Continuation-passing style

CPS conversion (with the more specific type for the auxilliary function, but this time making it a local definition) gives:

$$\begin{aligned}
&\mathit{reverse}_2 : \mathit{List}\ a \to \mathit{List}\ a \\
&\mathit{reverse}_2\ \mathit{xs} = \mathit{reverse}_2'\ \mathit{xs}\ \mathit{id}\ \ \mathbf{where} \\
&\quad \mathit{reverse}_2' : \mathit{List}\ a \to (\mathit{List}\ a \to \mathit{List}\ a) \to \mathit{List}\ a \\
&\quad \mathit{reverse}_2'\ [\,]\ k \qquad\ = k\ [\,] \\
&\quad \mathit{reverse}_2'\ (x :: \mathit{xs})\ k = \mathit{reverse}_2'\ \mathit{xs}\ (\lambda \mathit{zs} \Rightarrow k\ (\mathit{zs} +\!\!+ [x]))
\end{aligned}$$

## 3.2 Defunctionalization

The continuations are all functions of the form $(+\!\!+[a])\cdot(+\!\!+[b])\cdots(+\!\!+[c])$, compositions of functions that each append a single element. Defunctionalization picks a first-order encoding of such higher-order values; which encoding shall we pick? The most natural thing to try is to represent the composition as the list $[a, b, \ldots, c]$ of appendees. But then the abstraction function, which we need in the base case of the auxilliary function, is list reversal again! That's no help.

## 3.3 Associativity

The next most natural thing is to represent the composition $(+\!\!+[a])\cdot(+\!\!+[b])\cdots(+\!\!+[c])$ as the list $[c, \ldots, b, a]$ of appendees in reverse order—after all, the abstraction function then no longer needs to reverse anything:

$$\begin{aligned}
&\mathit{revabs}_3 : \mathit{List}\ a \to (\mathit{List}\ a \to \mathit{List}\ a) \\
&\mathit{revabs}_3\ \mathit{ys} = \lambda \mathit{zs} \Rightarrow \mathit{zs} +\!\!+ \mathit{ys}
\end{aligned}$$

Is this more helpful? Let's work through the development. We start with the specification

$$reverse'_3\ xs\ k = reverse'_2\ xs\ (revabs_3\ k) \qquad \text{-- specification of } reverse'_3$$

and calculate for the base case of the auxilliary function:

$$\begin{aligned}
& reverse'_3\ [\,]\ ys \\
= {} & [\![ \text{ specification of } reverse'_3\ ]\!] \\
& reverse'_2\ [\,]\ (revabs_3\ ys) \\
= {} & [\![ \text{ definition of } reverse'_2\ ]\!] \\
& revabs_3\ ys\ [\,] \\
= {} & [\![ \text{ definition of } revabs_3\ ]\!] \\
& [\,] +\!\!+ ys \\
= {} & [\![ \text{ definition of } +\!\!+\ ]\!] \\
& ys
\end{aligned}$$

and for the inductive step:

$$\begin{aligned}
& reverse'_3\ (x :: xs)\ ys \\
= {} & [\![ \text{ specification of } reverse'_3\ ]\!] \\
& reverse'_2\ (x :: xs)\ (revabs_3\ ys) \\
= {} & [\![ \text{ definition of } reverse'_2\ ]\!] \\
& reverse'_2\ xs\ (\lambda zs \Rightarrow revabs_3\ ys\ (zs +\!\!+ [x])) \\
= {} & [\![ \text{ definition of } revabs_3\ ]\!] \\
& reverse'_2\ xs\ (\lambda zs \Rightarrow (zs +\!\!+ [x]) +\!\!+ ys) \\
= {} & [\![ \text{ associativity of } +\!\!+\text{, and definition }\ ]\!] \\
& reverse'_2\ xs\ (\lambda zs \Rightarrow zs +\!\!+ (x :: ys)) \\
= {} & [\![ \text{ definition of } revabs_3\ ]\!] \\
& reverse'_2\ xs\ (revabs_3\ (x :: ys)) \\
= {} & [\![ \text{ specification of } reverse'_3\ ]\!] \\
& reverse'_3\ xs\ (x :: ys)
\end{aligned}$$

We end up with the well-known linear-time fast reverse function using an accumulator:

$$\begin{aligned}
& reverse_3 : List\ a \to List\ a \\
& reverse_3\ xs = reverse'_3\ xs\ [\,]\ \ \mathbf{where} \\
& \quad reverse'_3 : List\ a \to List\ a \to List\ a \\
& \quad reverse'_3\ [\,]\ k \qquad\quad = k \\
& \quad reverse'_3\ (x :: xs)\ k = reverse'_3\ xs\ (x :: k)
\end{aligned}$$

But we only get there by exploiting associativity of $+\!\!+$; without that, there is no productive defunctionalization of the CPS version of the program in the first place.

# 4 Flattening trees

Let's now turn our attention to non-linear data; in particular, binary trees:

$$\textbf{data}\ \mathit{Tree}\ a = \mathit{Tip}\ a \mid \mathit{Bin}\ (\mathit{Tree}\ a)\ (\mathit{Tree}\ a)$$

Here is the obvious, recursive, direct-style function for flattening a tree to a list:

$$\begin{aligned}&\mathit{flatten} : \mathit{Tree}\ a \to \mathit{List}\ a\\&\mathit{flatten}\ (\mathit{Tip}\ x) \quad = [x]\\&\mathit{flatten}\ (\mathit{Bin}\ t\ u) = \mathit{flatten}\ t \mathbin{+\!\!+} \mathit{flatten}\ u\end{aligned}$$

Note that, like *reverse*, this is again quadratic-time, because of the nested applications of $+\!\!+$; and again, like *reverse*, there is a linear-time implementation obtained via an accumulating parameter. We might suspect that CPS and defunctionalization will again take us there. Let's see.

## 4.1 Continuation-passing style

Following our nose with CPS conversion gives:

$$\begin{aligned}&\mathit{flatten}_2 : \mathit{Tree}\ a \to \mathit{List}\ a\\&\mathit{flatten}_2\ t = \mathit{flatten}_2'\ t\ \mathit{id}\ \ \textbf{where}\\&\quad \mathit{flatten}_2' : \mathit{Tree}\ a \to (\mathit{List}\ a \to \mathit{List}\ a) \to \mathit{List}\ a\\&\quad \mathit{flatten}_2'\ (\mathit{Tip}\ x)\ k \quad = k\ [x]\\&\quad \mathit{flatten}_2'\ (\mathit{Bin}\ t\ u)\ k = \mathit{flatten}_2'\ t\ (\lambda xs \Rightarrow \mathit{flatten}_2'\ u\ (\lambda ys \Rightarrow k\ (xs \mathbin{+\!\!+} ys)))\end{aligned}$$

Note that because binary trees are a non-linear datatype, there a choice to be made when using CPS to linearize the traversal: left-to-right, or right-to-left? This definition is left-to-right, visiting left child $t$ before right child $u$ (we will revisit this choice later).

## 4.2 Defunctionalization

It is perhaps no longer immediately obvious how to defunctionalize these continuations; what is the general form that they take? We need not wait for inspiration to strike: we can proceed methodically, following the path laid by Reynolds and explored by Danvy. There are three places where continuations are constructed, for the three call sites to $\mathit{flatten}_2'$:

- $\mathit{id}$ -- no free variables
- $\lambda ys \Rightarrow k\ (xs \mathbin{+\!\!+} ys)$ -- free variables $xs, k$
- $\lambda xs \Rightarrow \mathit{flatten}_2'\ u\ (\lambda ys \Rightarrow k\ (xs \mathbin{+\!\!+} ys))$ -- free variables $u, k$

So we will need a datatype with three different constructors. (Compare this with $\mathit{fact}_3$, which constructed continuations at the two call sites to the auxilliary function $\mathit{fact}_3'$, so used a datatype with two constructors, namely lists; and similarly for $\mathit{reverse}_2$.)

The last two of the three call sites to $\mathit{flatten}_2'$ construct continuations with free variables, representing the context of the recursive call; the corresponding constructors will be *closures*, taking arguments to capture values for those free variables. We therefore introduce the following representation:

$$\begin{aligned}\textbf{data}\ \mathit{FlatCont}_3\ a = {} & \mathit{FlatRoot}_3 \\ \mid {} & \mathit{FlatLeftTree}_3\ (\mathit{Tree}\ a)\ (\mathit{FlatCont}_3\ a) \\ \mid {} & \mathit{FlatRightList}_3\ (\mathit{List}\ a)\ (\mathit{FlatCont}_3\ a)\end{aligned}$$

The idea is that $\mathit{FlatRoot}_3$ is the continuation used at the root of the tree, $\mathit{FlatLeftTree}_3$ for the recursive call on a left child (whose context includes the right sibling, another tree), and $\mathit{FlatRightList}_3$ for the recursive call on a right child (whose context includes the flattening of the left sibling, a list). The abstraction function is typed

$$\mathit{flatabs}_3 : \mathit{FlatCont}_3\ a \rightarrow (\mathit{List}\ a \rightarrow \mathit{List}\ a)$$

and the corresponding auxilliary function is specified by

$$\mathit{flatten}'_3\ t\ k = \mathit{flatten}'_2\ t\ (\mathit{flatabs}_3\ k) \qquad \text{-- specification of } \mathit{flatten}'_3$$

Working through the data refinement yields the following implementation:

$$\begin{aligned}
&\mathit{flatten}_3 : \mathit{Tree}\ a \rightarrow \mathit{List}\ a \\
&\mathit{flatten}_3\ t = \mathit{flatten}'_3\ t\ \mathit{FlatRoot}_3\ \ \textbf{where} \\
&\quad \textbf{mutual} \\
&\qquad \mathit{flatten}'_3 : \mathit{Tree}\ a \rightarrow \mathit{FlatCont}_3\ a \rightarrow \mathit{List}\ a \\
&\qquad \mathit{flatten}'_3\ (\mathit{Tip}\ x)\ k \quad = \mathit{flatabs}_3\ k\ [x] \\
&\qquad \mathit{flatten}'_3\ (\mathit{Bin}\ t\ u)\ k = \mathit{flatten}'_3\ t\ (\mathit{FlatLeftTree}_3\ u\ k) \\
&\qquad \mathit{flatabs}_3 : \mathit{FlatCont}_3\ a \rightarrow (\mathit{List}\ a \rightarrow \mathit{List}\ a) \\
&\qquad \mathit{flatabs}_3\ \mathit{FlatRoot}_3 \qquad\qquad\quad = \mathit{id} \\
&\qquad \mathit{flatabs}_3\ (\mathit{FlatLeftTree}_3\ u\ k) \quad = \lambda xs \Rightarrow \mathit{flatten}'_3\ u\ (\mathit{FlatRightList}_3\ xs\ k) \\
&\qquad \mathit{flatabs}_3\ (\mathit{FlatRightList}_3\ xs\ k) = \lambda ys \Rightarrow \mathit{flatabs}_3\ k\ (xs \mathbin{+\!\!+} ys)
\end{aligned}$$

Intuitively, the defunctionalized continuation is a stack of 'context frames', each of which is either a tree (a right sibling, the additional context when descending into a left child) or a list (the flattening of a left sibling, the additional context when descending into a right child).

### 4.3 Defunctionalization again

By happy accident, we need not have defined a new algebraic datatype for the defunctionalized continuations, since $\mathit{FlatCont}_3\ a$ happens to be isomorphic to a sequence of choices:

$$\begin{aligned}&\mathit{FlatCont}_4 : \mathit{Type} \rightarrow \mathit{Type} \\ &\mathit{FlatCont}_4\ a = \mathit{List}\ (\mathit{Either}\ (\mathit{Tree}\ a)\ (\mathit{List}\ a))\end{aligned}$$

(here, the sum type $\mathit{Either}\ X\ Y$ has values of the form $\mathit{Left}\ x$ for $x : X$ and $\mathit{Right}\ y$ for $y : Y$). Using this representation instead gives the equivalent program:

$$
\begin{array}{l}
\mathit{flatten}_4 : \mathit{Tree}\ a \to \mathit{List}\ a \\
\mathit{flatten}_4\ t = \mathit{flatten}'_4\ t\ [\,]\ \ \textbf{where} \\
\quad \textbf{mutual} \\
\qquad \mathit{flatten}'_4 : \mathit{Tree}\ a \to \mathit{FlatCont}_4\ a \to \mathit{List}\ a \\
\qquad \mathit{flatten}'_4\ (\mathit{Tip}\ x)\ k \quad = \mathit{flatabs}_4\ k\ [x] \\
\qquad \mathit{flatten}'_4\ (\mathit{Bin}\ t\ u)\ k = \mathit{flatten}'_4\ t\ (\mathit{Left}\ u :: k) \\
\\
\qquad \mathit{flatabs}_4 : \mathit{FlatCont}_4\ a \to (\mathit{List}\ a \to \mathit{List}\ a) \\
\qquad \mathit{flatabs}_4\ [\,] \qquad\qquad\quad = \mathit{id} \\
\qquad \mathit{flatabs}_4\ (\mathit{Left}\ u :: k) \quad\ = \lambda xs \Rightarrow \mathit{flatten}'_4\ u\ (\mathit{Right}\ xs :: k) \\
\qquad \mathit{flatabs}_4\ (\mathit{Right}\ xs :: k) = \lambda ys \Rightarrow \mathit{flatabs}_4\ k\ (xs +\!\!+\, ys)
\end{array}
$$

With built-in lists, it is easier to see that the defunctionalized continuation behaves as a stack. However, the naming convention is a bit awkward with the built-in sum type: a right child $u$ is in the *Left* side of the sum, because it is the context for descending into its left sibling.

Nevertheless, this is not the linear-time flattening function we were hoping for: it's still quadratic, because every flattening $xs$ of a left sibling is retraversed to append the flattening $ys$ of its right sibling (in the *Right* case of $\mathit{flatabs}_4$). Where did we go wrong?

### 4.4 Continuation-passing style again

As noted above, there was actually an arbitrary decision made earlier, not present in the direct-style program *flatten* and forced upon us by the CPS conversion to $\mathit{flatten}_2$. The direct-style program is agnostic about whether left children are visited before or after right children; that is determined by the evaluation semantics of the defining language, and the definition of $+\!\!+$. The CPS program makes explicit that the left child is visited before the right. Making this order of evaluation explicit is precisely why Reynolds used CPS for his definitional interpreters; and 'left before right' turns out to be precisely the wrong thing to do for tree flattening. Here is the other CPS version, visiting the right child before the left (but still delivering the tips in left to right order):

$$
\begin{array}{l}
\mathit{flatten}_5 : \mathit{Tree}\ a \to \mathit{List}\ a \\
\mathit{flatten}_5\ t = \mathit{flatten}'_5\ t\ \mathit{id}\ \ \textbf{where} \\
\quad \mathit{flatten}'_5 : \mathit{Tree}\ a \to (\mathit{List}\ a \to \mathit{List}\ a) \to \mathit{List}\ a \\
\quad \mathit{flatten}'_5\ (\mathit{Tip}\ x)\ k \quad = k\ [x] \\
\quad \mathit{flatten}'_5\ (\mathit{Bin}\ t\ u)\ k = \mathit{flatten}'_5\ u\ (\lambda ys \Rightarrow \mathit{flatten}'_5\ t\ (\lambda xs \Rightarrow k\ (xs +\!\!+\, ys)))
\end{array}
$$

These continuations defunctionalize in the mirror image of what happened earlier: the additional context for a right child $u$ is its left sibling $t$, and the context for a left child $t$ is the flattening $ys$ of its right sibling. So we introduce a new representation of defunctionalized continuations:

$$
\begin{array}{l}
\mathit{FlatCont}_6 : \mathit{Type} \to \mathit{Type} \\
\mathit{FlatCont}_6\ a = \mathit{List}\ (\mathit{Either}\ (\mathit{List}\ a)\ (\mathit{Tree}\ a))
\end{array}
$$

Using this representation instead gives the following defunctionalized right-to-left traversal:

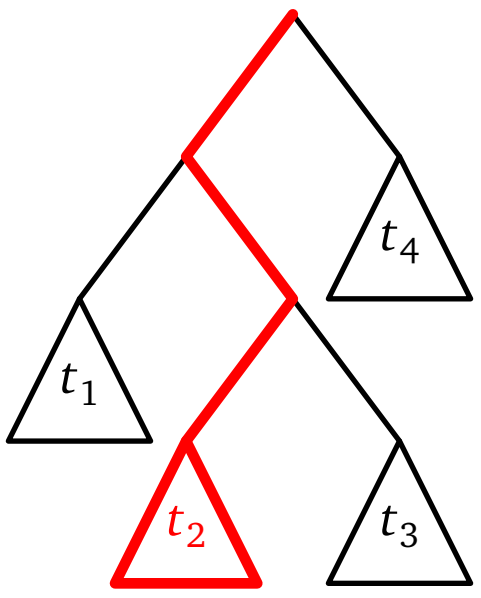

**Figure 1** A tree to be flattened, with subtree $t_2$ highlighted

$$
\begin{aligned}
&\mathit{flatten}_6 : \mathit{Tree}\ a \rightarrow \mathit{List}\ a \\
&\mathit{flatten}_6\ t = \mathit{flatten}'_6\ t\ [\,]\ \textbf{where} \\
&\quad \textbf{mutual} \\
&\qquad \mathit{flatten}'_6 : \mathit{Tree}\ a \rightarrow \mathit{FlatCont}_6\ a \rightarrow \mathit{List}\ a \\
&\qquad \mathit{flatten}'_6\ (\mathit{Tip}\ x)\ k \quad = \mathit{flatabs}_6\ k\ [x] \\
&\qquad \mathit{flatten}'_6\ (\mathit{Bin}\ t\ u)\ k = \mathit{flatten}'_6\ u\ (\mathit{Right}\ t :: k) \\
&\qquad \mathit{flatabs}_6 : \mathit{FlatCont}_6\ a \rightarrow (\mathit{List}\ a \rightarrow \mathit{List}\ a) \\
&\qquad \mathit{flatabs}_6\ [\,] \qquad\qquad = \mathit{id} \\
&\qquad \mathit{flatabs}_6\ (\mathit{Left}\ ys :: k)\ \ = \lambda xs \Rightarrow \mathit{flatabs}_6\ k\ (xs +\!\!+ ys) \\
&\qquad \mathit{flatabs}_6\ (\mathit{Right}\ t :: k) = \lambda ys \Rightarrow \mathit{flatten}'_6\ t\ (\mathit{Left}\ ys :: k)
\end{aligned}
$$

Sadly, this is still quadratic-time: in the *Left* case of $\mathit{flatabs}_6$, the constructed flattening *xs* is retraversed to append *ys*. More work is required to get linear time. And not surprisingly, that additional work involves associativity.

### 4.5 Associativity

Consider applying $\mathit{flatten}_6$ to the tree shown in figure 1, via

$$xs = \mathit{flatten}_6\ (\mathit{Bin}\ (\mathit{Bin}\ t_1\ (\mathit{Bin}\ t_2\ t_3))\ t_4)$$

where $t_1, t_2, t_3, t_4$ are the four subtrees; and let $xs_i = \mathit{flatten}\ t_i$ for $i = 1, 2, 3, 4$. The figure highlights the left–right–left path leading from the root to subtree $t_2$. The context when visiting subtree $t_2$ is the continuation $[\mathit{Left}\ xs_3, \mathit{Right}\ t_1, \mathit{Left}\ xs_4]$; that is,

$$xs = \mathit{flatten}'_6\ t_2\ [\mathit{Left}\ xs_3, \mathit{Right}\ t_1, \mathit{Left}\ xs_4]$$

Having obtained $xs_2 = \mathit{flatten}\ t_2$, the continuation is applied, leading eventually to

$$xs = (\mathit{flatten}\ t_1 +\!\!+ (xs_2 +\!\!+ xs_3)) +\!\!+ xs_4$$

This reveals that the particular interleaving of *Left* and *Right* frames in the continuation is irrelevant: *Left* frames $xs_3, xs_4$ get appended, *Right* frames $t_1$ get prepended, and—by associativity of $+\!\!+$!—these two operations commute. So we could just as well separate the *Left* frames from the *Right* frames:

$$\mathit{flatabs}_6\ k = \mathit{flatabs}_6\ (\mathit{map}\ \mathit{Left}\ (\mathit{lefts}\ k) +\!\!+ \mathit{map}\ \mathit{Right}\ (\mathit{rights}\ k))$$

where $lefts : List\ (Either\ a\ b) \to List\ a$ and $rights : List\ (Either\ a\ b) \to List\ b$ extract the *Left*- and *Right*-tagged elements respectively of a list of *Either*s. This in turn suggests another data refinement: we can discard the interleaving information, and represent the defunctionalized continuation instead as an uninterleaved pair of lists, $(List\ (List\ a), List\ (Tree\ a))$, consisting respectively of the data to be appended and that to be prepended.

But we can (and should) go further: all the *Left*-tagged lists are eventually going to be concatenated, so we might as well concatenate them into a single list as we go. That is,

$$flatabs_6\ k = flatabs_6\ ([Left\ (concat\ (lefts\ k))] +\!\!+\ map\ Right\ (rights\ k))$$

This suggests yet another data refinement, to represent the defunctionalized continuation as a pair consisting of a single list of elements (to be appended) and a list of trees (whose flattenings will be prepended). We therefore introduce the new representation of defunctionalized continuations

$$\begin{aligned} &FlatCont_7 : Type \to Type \\ &FlatCont_7\ a = (List\ a, List\ (Tree\ a)) \end{aligned}$$

with abstraction function specified by

$$flatabs_7\ (ys, ts)\ xs = concat\ (map\ flatten\ (reverse\ ts)) +\!\!+\ xs +\!\!+\ ys \qquad \text{-- specification of } flatabs_7$$

Applying this data refinement leads to the following program:

$$\begin{aligned} &flatten_7 : Tree\ a \to List\ a \\ &flatten_7\ t = flatten'_7\ t\ ([\,],[\,])\ \ \mathbf{where} \\ &\quad \mathbf{mutual} \\ &\quad\quad flatten'_7 : Tree\ a \to FlatCont_7\ a \to List\ a \\ &\quad\quad flatten'_7\ (Tip\ x)\ (ys, ts) \quad = flatabs_7\ (ys, ts)\ [x] \\ &\quad\quad flatten'_7\ (Bin\ t\ u)\ (ys, ts) = flatten'_7\ u\ (ys, t :: ts) \\ &\quad\quad flatabs_7 : FlatCont_7\ a \to (List\ a \to List\ a) \\ &\quad\quad flatabs_7\ (ys, [\,]) \qquad = \lambda xs \Rightarrow xs +\!\!+\ ys \\ &\quad\quad flatabs_7\ (ys, t :: ts) = \lambda xs \Rightarrow flatten'_7\ t\ (xs +\!\!+\ ys, ts) \end{aligned}$$

This still contains two occurrences of $+\!\!+$, in the two clauses for $flatabs_7$. But note that $flatabs_7$ is only applied in a single place, and then only to a singleton list; so those $+\!\!+$s take constant time. This program therefore takes linear time overall; and the transformations that brought us there depended crucially on associativity of $+\!\!+$.

This is still not quite the program you might have written from first principles, but it is only one final data refinement away. Instead of the call $flatten'_7\ t\ (ys, ts)$ with an isolated 'tree in focus' $t$ and a possibly empty stack $ts$ of postponed trees, we push $t$ onto $ts$ to make a non-empty stack $t :: ts$ of trees:

$$\begin{aligned} &flatten_8 : Tree\ a \to List\ a \\ &flatten_8\ t = flatten'_8\ ([\,],[t])\ \ \mathbf{where} \end{aligned}$$

**mutual**

$$
\begin{array}{l}
\mathit{flatten}'_8 : (\mathit{List}\ a, \mathit{List}\ (\mathit{Tree}\ a)) \to \mathit{List}\ a \\
\mathit{flatten}'_8\ (ys, \mathit{Tip}\ x :: ts) \quad = \mathit{flatabs}_8\ (ys, ts)\ [x] \\
\mathit{flatten}'_8\ (ys, \mathit{Bin}\ t\ u :: ts) = \mathit{flatten}'_8\ (ys, u :: t :: ts) \\
\mathit{flatabs}_8 : (\mathit{List}\ a, \mathit{List}\ (\mathit{Tree}\ a)) \to (\mathit{List}\ a \to \mathit{List}\ a) \\
\mathit{flatabs}_8\ (ys, [\,]) \quad = \lambda xs \Rightarrow xs +\!\!+ ys \\
\mathit{flatabs}_8\ (ys, t :: ts) = \lambda xs \Rightarrow \mathit{flatten}'_8\ (xs +\!\!+ ys, t :: ts)
\end{array}
$$

There is only one call site for $\mathit{flatabs}_8$; if we make a case analysis there on whether *ts* is empty, we can inline $\mathit{flatabs}_8$ and eliminate the additional definition. We also take the opportunity to curry the auxilliary function, and to flip its arguments:

$$
\begin{array}{l}
\mathit{flatten}_9 : \mathit{Tree}\ a \to \mathit{List}\ a \\
\mathit{flatten}_9\ t = \mathit{flatten}'_9\ [t]\ [\,]\ \ \textbf{where} \\
\quad \mathit{flatten}'_9 : \mathit{List}\ (\mathit{Tree}\ a) \to \mathit{List}\ a \to \mathit{List}\ a \\
\quad \mathit{flatten}'_9\ [\mathit{Tip}\ x] \qquad ys = x :: ys \\
\quad \mathit{flatten}'_9\ (\mathit{Tip}\ x :: ts) \quad ys = \mathit{flatten}'_9\ ts\ (x :: ys) \\
\quad \mathit{flatten}'_9\ (\mathit{Bin}\ t\ u :: ts)\ ys = \mathit{flatten}'_9\ (u :: t :: ts)\ ys
\end{array}
$$

This program operates on a non-empty stack of trees, maintaining an accumulating parameter for the flattening so far, which is constructed from right to left. It is tail-recursive, first-order, and linear-time. This *is* the program you would write from first principles: a model application of an accumulating parameter to tree flattening.

## 5 Interpreters and compilers

Reynolds' original motivation [28] for investigating CPS and defunctionalization was to elucidate the derivation of abstract machines from language interpreters. Associativity also shows up there—in a way not exploited by Reynolds himself, nor by the extensive subsequent elaboration by Danvy and colleagues [13, 2, 1, 3, 4, 12], but it is central to the approach taken by Wand [31, 32]. However, in contrast to our earlier examples using factorial, reverse, and flattening, associativity does not arise from the problem domain—that is, the operators of the language being implemented—but from (a generalized form of) sequential composition of subprograms. Our point here is to remind ourselves of Wand's approach, which seems to have become lost over the intervening four decades.

For these purposes, it suffices to take an extremely simple language—one of arithmetic expressions constructed from numeric literals and addition. Hutton [18] shows that this same language serves as a basis for discussing many aspects of language semantics and implementation. But because we are interested specifically in associativity, we diverge from Hutton's language to one of literals and (non-associative) *subtraction*; we do not want to confuse associativity of sequential composition in the machine implementation with associativity of addition in the source language.

$$\textbf{data}\ \mathit{Expr} = \mathit{Lit}\ \mathit{Integer} \mid \mathit{Diff}\ \mathit{Expr}\ \mathit{Expr}$$

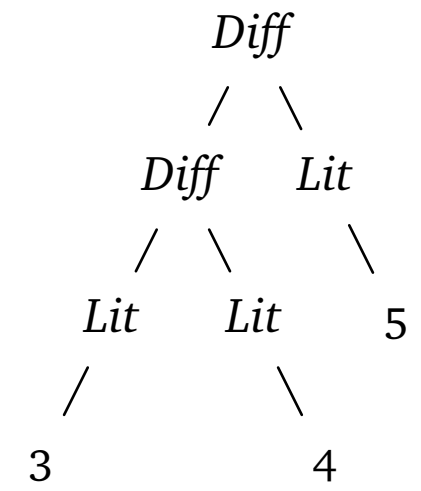

**Figure 2** The expression *Diff* (*Diff* (*Lit* 3) (*Lit* 4)) (*Lit* 5)

Here is the straightforward, direct-style, recursive evaluator for this language:

$$\begin{aligned}
&\mathit{eval} : \mathit{Expr} \to \mathit{Integer}\\
&\mathit{eval}\ (\mathit{Lit}\ n) \quad\ \ = n\\
&\mathit{eval}\ (\mathit{Diff}\ e\ e') = \mathit{eval}\ e - \mathit{eval}\ e'
\end{aligned}$$

so with

$$\mathit{expr} = \mathit{Diff}\ (\mathit{Diff}\ (\mathit{Lit}\ 3)\ (\mathit{Lit}\ 4))\ (\mathit{Lit}\ 5)$$

as shown in figure 2 we have $\mathit{eval}\ \mathit{expr} = (3-4)-5 = -6$.

### 5.1 CPS and defunctionalization

If we follow the same process as for tree flattening, we obtain a tail-recursive evaluator via CPS:

$$\begin{aligned}
&\mathit{eval}_2 : \mathit{Expr} \to \mathit{Integer}\\
&\mathit{eval}_2\ e = \mathit{eval}'_2\ e\ \mathit{id}\ \ \textbf{where}\\
&\quad \mathit{eval}'_2 : \mathit{Expr} \to (\mathit{Integer} \to \mathit{Integer}) \to \mathit{Integer}\\
&\quad \mathit{eval}'_2\ (\mathit{Lit}\ n)\ k \quad\ \ = k\ n\\
&\quad \mathit{eval}'_2\ (\mathit{Diff}\ e\ e')\ k = \mathit{eval}'_2\ e\ (\lambda m \Rightarrow \mathit{eval}'_2\ e'\ (\lambda n \Rightarrow k\ (m-n)))
\end{aligned}$$

and a tail-recursive, first-order evaluator via defunctionalization:

$$\begin{aligned}
&\textbf{data}\ \mathit{EvalFrame}_3 = \mathit{EvalLeftExpr}_3\ \mathit{Expr} \mid \mathit{EvalRightValue}_3\ \mathit{Integer}\\
&\mathit{EvalCont}_3 : \mathit{Type}\\
&\mathit{EvalCont}_3 = \mathit{List}\ \mathit{EvalFrame}_3\\
&\mathit{eval}_3 : \mathit{Expr} \to \mathit{Integer}\\
&\mathit{eval}_3\ e = \mathit{eval}'_3\ e\ [\,]\ \ \textbf{where}\\
&\quad \textbf{mutual}\\
&\qquad \mathit{eval}'_3 : \mathit{Expr} \to \mathit{EvalCont}_3 \to \mathit{Integer}\\
&\qquad \mathit{eval}'_3\ (\mathit{Lit}\ n) \qquad k = \mathit{evalabs}_3\ k\ n\\
&\qquad \mathit{eval}'_3\ (\mathit{Diff}\ e\ e')\ k = \mathit{eval}'_3\ e\ (\mathit{EvalLeftExpr}_3\ e' :: k)\\
&\qquad \mathit{evalabs}_3 : \mathit{EvalCont}_3 \to (\mathit{Integer} \to \mathit{Integer})\\
&\qquad \mathit{evalabs}_3\ [\,] \qquad\qquad\qquad\qquad\quad\ n\ \ = n\\
&\qquad \mathit{evalabs}_3\ (\mathit{EvalLeftExpr}_3\ e' :: k) \quad\ \ m = \mathit{eval}'_3\ e'\ (\mathit{EvalRightValue}_3\ m :: k)\\
&\qquad \mathit{evalabs}_3\ (\mathit{EvalRightValue}_3\ m :: k)\ n\ \ = \mathit{evalabs}_3\ k\ (m-n)
\end{aligned}$$

The defunctionalized continuations $EvalCont_3$ are stacks of frames, with either unevaluated expressions (the right siblings of left visitees) or integers (the evaluations of left siblings of right visitees); we have defined a new datatype $EvalFrame_3$ to allow more specific names than with plain *Either Expr Integer*.

Now $eval_3$ is readily seen as an abstract machine for evaluating expressions: evaluating a *Diff* pushes the right child on the stack then moves down to the left; when eventually a result is obtained, the next task is popped off the stack; if this result was from a left child, its right sibling is popped off the stack to visit next and the result pushed in its place; and if the result was from a right child, the corresponding left result is popped from the stack, the difference computed, and the remainder of the stack used; when the stack is empty, the result obtained is the final result.

### 5.2 Compilation

However, $eval_3$ cannot be seen as a *compiler* for expressions: the abstract machine manipulates unevaluated expressions 'at run-time' on the stack, whereas a compiler should have eliminated these. Nevertheless, there is a well-known stack-based compilation scheme for such expression languages. There are instructions to push a value onto the stack, and to replace the top two elements of the stack with their difference:

$$
\begin{array}{l}
\mathbf{data}\ \mathit{Instr} = \mathit{PushI}\ \mathit{Integer} \mid \mathit{SubI} \\
\mathit{Prog}_4 : \mathit{Type} \\
\mathit{Prog}_4 = \mathit{List}\ \mathit{Instr}
\end{array}
$$

Program execution is a left fold over the instruction sequence, inducing a stack transformation (a function from stacks to stacks):

$$
\begin{array}{l}
\mathit{exec}_4 : \mathit{Prog}_4 \to \mathit{List}\ \mathit{Integer} \to \mathit{List}\ \mathit{Integer} \\
\mathit{exec}_4\ p\ s = \mathit{foldl}\ \mathit{step}\ s\ p\ \ \mathbf{where} \\
\quad \mathit{step}\ \mathit{ns} \qquad\qquad\quad (\mathit{PushI}\ n) = n :: \mathit{ns} \\
\quad \mathit{step}\ (n :: m :: \mathit{ns})\ \mathit{SubI} \qquad = (m - n) :: \mathit{ns} \qquad\qquad \text{-- note reversal of arguments}
\end{array}
$$

Compilation is a traversal of the expression tree:

$$
\begin{array}{l}
\mathit{compile}_4 : \mathit{Expr} \to \mathit{Prog}_4 \\
\mathit{compile}_4\ (\mathit{Lit}\ n) \qquad = [\mathit{PushI}\ n] \\
\mathit{compile}_4\ (\mathit{Diff}\ e\ e') = \mathit{compile}_4\ e +\!\!+\ \mathit{compile}_4\ e' +\!\!+ [\mathit{SubI}]
\end{array}
$$

For example,

$$\mathit{compile}_4\ \mathit{expr} = [\mathit{PushI}\ 3, \mathit{PushI}\ 4, \mathit{SubI}, \mathit{PushI}\ 5, \mathit{SubI}]$$

In the special case of a program of the form $\mathit{compile}_4\ e$, execution consists of consuming an initial stack (which is not inspected) and producing a final stack with the evaluation of $e$ pushed on top:

$$\mathit{exec}_4\ (\mathit{compile}_4\ e)\ s = \mathit{eval}\ e :: s$$

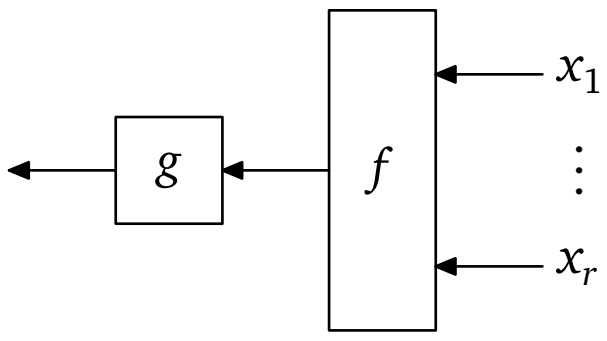

**Figure 3** Generalized composition $b^r\, g\, f$

In particular, an expression may be evaluated by compiling and executing a program on the initially empty stack, then extracting the sole element of the final stack:

$$eval_4 : Expr \rightarrow Integer$$
$$eval_4\ e = \mathbf{case}\ exec_4\ (compile_4\ e)\ [\,]\ \mathbf{of}\ [\,n\,] \Rightarrow n$$

This compilation scheme entails an abstract machine, and it evidently involves stacks; so it seems like it ought to be obtainable using CPS and defunctionalization. Can we get there following the same process as with the other examples? We can—but only if we exploit associativity.

### 5.3 Generalized composition

Wand [32] observes that the essence of the key transformation is associativity of a generalization of function composition. Consider the recursive case of the CPS interpreter:

$$eval_2'\ (Diff\ e\ e')\ k = eval_2'\ e\ (\lambda m \Rightarrow eval_2'\ e'\ (\lambda n \Rightarrow k\ (m - n)))$$

The right-hand side can be seen as routing one argument $k$ from the surrounding context into the continuation for the left child $e$, and routing two arguments $k, m$ into the continuation for the right child $e'$. To arrange this routing, Wand [32] therefore introduces the generalized composition operation $b^r$, such that

$$b^r\ g\ f\ x_1 \ldots x_r = g\ (f\ x_1 \ldots x_r)$$

as illustrated in figure 3 (although Wand writes $\mathsf{B}_r$ where we write $b^r$). Equivalently, by $\eta$-expansion,

$$\lambda x_1 \ldots x_r \Rightarrow b^r\ g\ f\ x_1 \ldots x_r = g\ (\lambda x_1 \ldots x_r \Rightarrow f\ x_1 \ldots x_r)$$

That is,

$$b^0 \quad g\ f = g\ f \qquad\qquad b^0 \quad = id$$
$$b^{r+1}\ g\ f = (b^r\ g) \cdot f \qquad\qquad b^{r+1} = (b^r\ g) \cdot (\cdot)$$

so that $b^1 = (\cdot)$, $b^2 = (\cdot) \cdot (\cdot)$, $b^3 = (\cdot) \cdot (\cdot) \cdot (\cdot)$— respectively Smullyan's 'bluebird', 'blackbird', and 'bunting' combinators [29] —and in general $b^r = (\cdot) \cdot \cdots \cdot (\cdot)$, the composition of $r$ instances of $(\cdot)$.

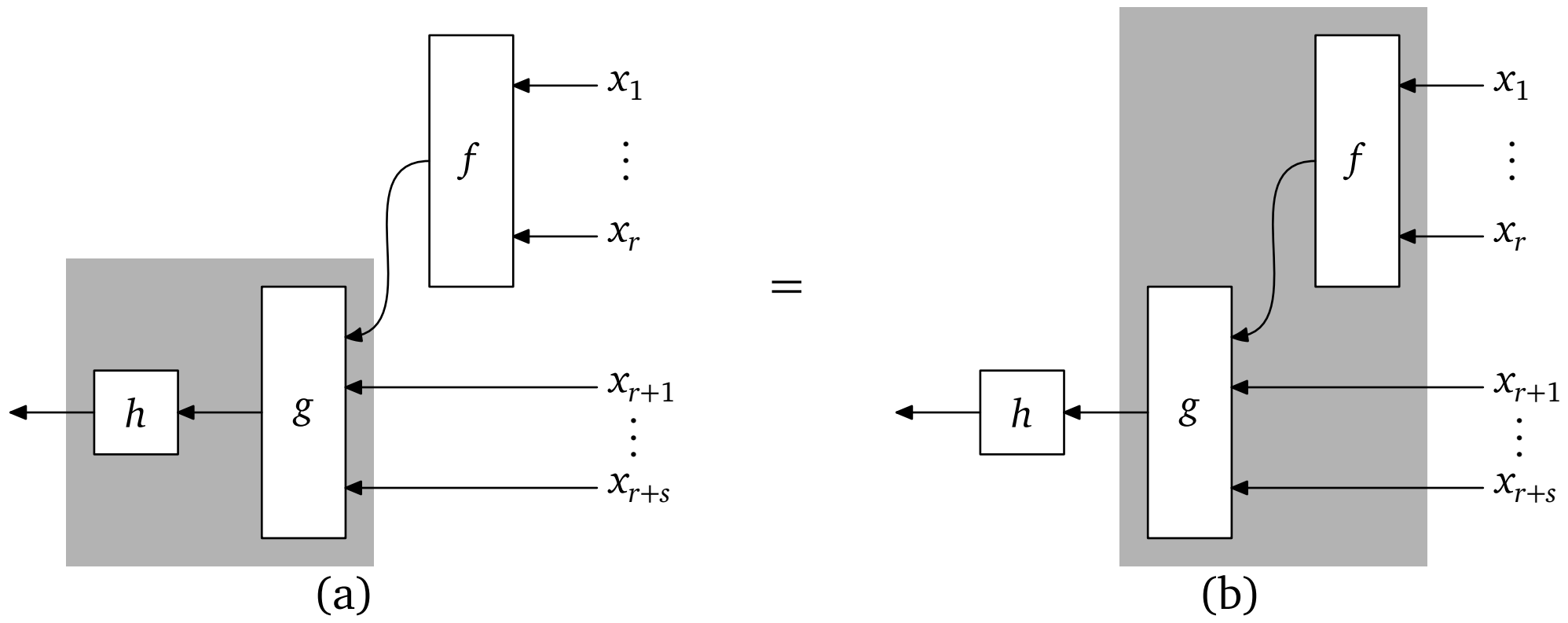

**Figure 4** Associativity of generalized composition: (a) $b^r\ (b^{s+1}\ h\ g)\ f$, (b) $b^{r+s}\ h\ (b^r\ g\ f)$

Using generalized composition, we have:

$$
\begin{aligned}
& eval'_2\ (Diff\ e\ e') \\
=\ & [\![ \text{ definition of } eval'_2\ ]\!] \\
& \lambda k \Rightarrow eval'_2\ e\ (\lambda m \Rightarrow eval'_2\ e'\ (\lambda n \Rightarrow k\ (m - n))) \\
=\ & [\![ \text{ since } \lambda k \Rightarrow g\ (f\ k) \text{ is } b^1\ g\ (\lambda k \Rightarrow f\ k)\ ]\!] \\
& b^1\ (eval'_2\ e)\ (\lambda k\ m \Rightarrow eval'_2\ e'\ (\lambda n \Rightarrow k\ (m - n))) \\
=\ & [\![ \text{ since } \lambda k\ m \Rightarrow g\ (f\ k\ m) \text{ is } b^2\ g\ (\lambda k\ m \Rightarrow f\ k\ m)\ ]\!] \\
& b^1\ (eval'_2\ e)\ (b^2\ (eval'_2\ e')\ (\lambda k\ m\ n \Rightarrow k\ (m - n)))
\end{aligned}
$$

Moreover, generalized composition is—of course!—(pseudo-)associative:

$$
\begin{aligned}
& (b^r\ (b^{s+1}\ h\ g)\ f)\ x_1 \ldots x_{r+s} \\
=\ & (b^{s+1}\ h\ g)\ (f\ x_1 \ldots x_r)\ x_{r+1} \ldots x_{r+s} \\
=\ & h\ (g\ (f\ x_1 \ldots x_r)\ x_{r+1} \ldots x_{r+s}) \\
=\ & h\ ((b^r\ g\ f)\ x_1 \ldots x_{r+s}) \\
=\ & (b^{r+s}\ h\ (b^r\ g\ f))\ x_1 \ldots x_{r+s}
\end{aligned}
$$

That is,

$$b^r\ (b^{s+1}\ h\ g)\ f = b^{r+s}\ h\ (b^r\ g\ f)$$

as illustrated in figure 4. We call this ‘pseudo-associativity’ rather than plain ‘associativity’ because the indices change under rebracketing. It is this associativity that is the crucial transformation allowing us to rotate tree-shaped code to obtain linear code, thereby eliminating expressions from the run-time artifacts, as we shall see.

### 5.4 Implementing generalized composition

The first few arities of generalized composition can be defined as follows:

$$
\begin{aligned}
& b^0 : (b \to c) \to b \to c \\
& b^0\ g\ f = g\ f \\
& b^1 : (b \to c) \to (a \to b) \to (a \to c) \\
& b^1\ g\ f = \lambda x \Rightarrow g\ (f\ x)
\end{aligned}
$$

$$b^2 : (b \to c) \to (a \to a' \to b) \to (a \to a' \to c)$$
$$b^2\, g\, f = \lambda x\, y \Rightarrow g\, (f\, x\, y)$$

However, the general case has a dependent type, depending on the arity of $f$. In order to accommodate different types for the different argument positions, we index the definition by a *list of types*. Thus, we introduce the type function $Arrow\ as\ b$ for list of types $as$ and type $b$:

$$\begin{array}{l} Arrow : List\ Type \to Type \to Type \\ Arrow\ [\,]\qquad b = b \\ Arrow\ (a :: as)\ b = a \to Arrow\ as\ b \end{array}$$

For example

$$Arrow\ [Char, Bool]\ String = Char \to Bool \to String$$

at arity 2. Then we can define generalized composition, by induction over the arity:

$$\begin{array}{l} b : \{as : List\ Type\} \to (b \to c) \to Arrow\ as\ b \to Arrow\ as\ c \\ b\ \{as = [\,]\}\quad g\, f = g\, f \\ b\ \{as = _ :: _\}\ g\, f = b\, g \cdot f \end{array}$$

Rewriting $eval'_2$ to use generalized composition yields a different implementation of the evaluator:

$$\begin{array}{l} eval_5 : Expr \to Integer \\ eval_5\ e = eval'_5\ e\ halt\ \ \textbf{where} \\ \quad eval'_5 : Expr \to (Integer \to Integer) \to Integer \\ \quad eval'_5\ (Lit\ n) \qquad = ret\ n \\ \quad eval'_5\ (Diff\ e\ e') = b^1\ (eval'_5\ e)\ (b^2\ (eval'_5\ e')\ sub) \end{array}$$

where for later convenience we have introduced three abbreviations:

$$\begin{array}{l} halt\ \ = id \\ ret\ n = \lambda k \Rightarrow k\ n \\ sub\ \ \ = \lambda k \Rightarrow \lambda m \Rightarrow \lambda n \Rightarrow k\ (m - n) \end{array}$$

### 5.5 Tree-shaped code

Note that $eval'_5$ is not tail-recursive any more, because of the imposition of the generalized compositions. Nevertheless, it suggests a data representation:

$$\begin{array}{lll} \multicolumn{3}{l}{\textbf{data}\ ExprRep_6 : List\ Type \to Type\ \ \textbf{where}} \\ \quad Ret_6 & : Integer \to & ExprRep_6\ [\,] \\ \quad Sub_6 & : & ExprRep_6\ [Integer, Integer] \\ \quad B^1_6 & : ExprRep_6\ [\,] \to ExprRep_6\ [Integer] \to & ExprRep_6\ [\,] \\ \quad B^2_6 & : ExprRep_6\ [\,] \to ExprRep_6\ [Integer, Integer] \to & ExprRep_6\ [Integer] \end{array}$$

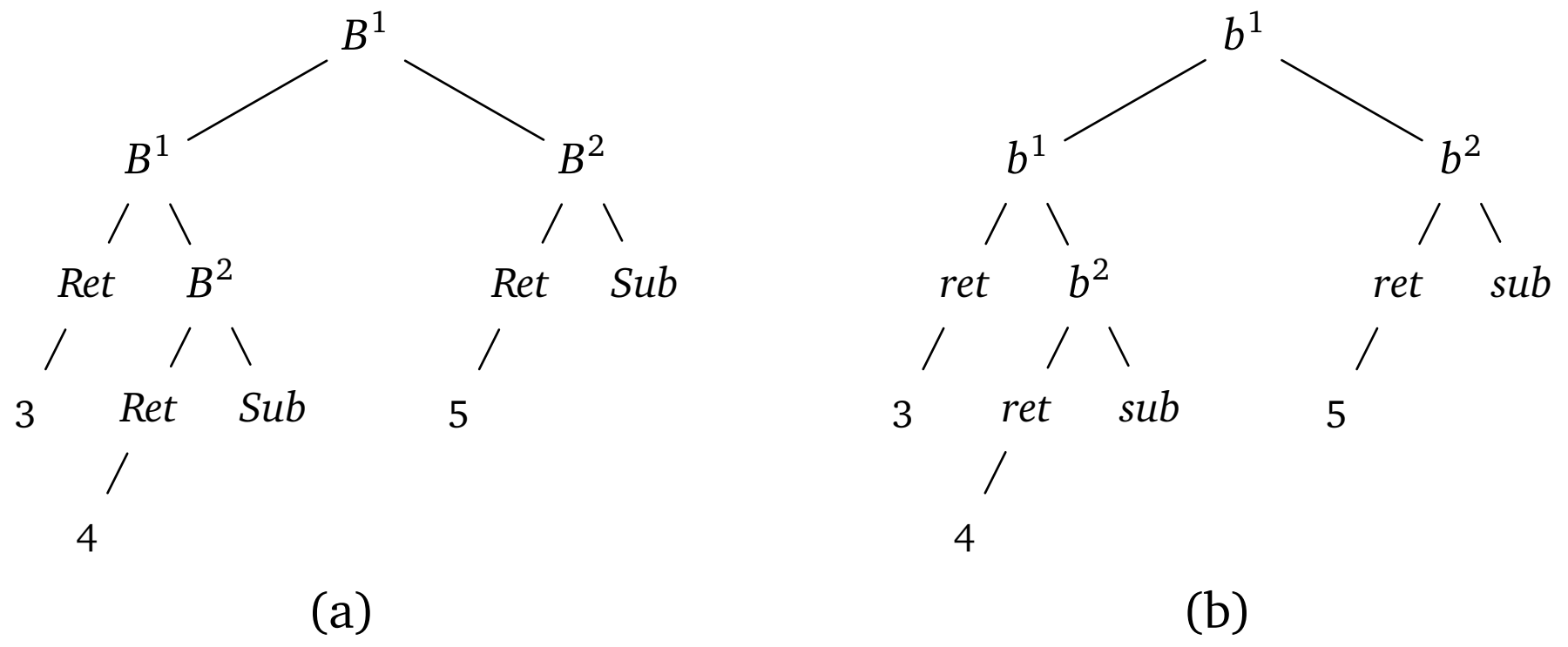

**Figure 5** (a) Tree-shaped code and (b) its interpretation

A data structure of type $ExprRep_6\ r$ represents an evaluation function of type $(Integer \to Integer) \to Arrow\ r\ Integer$. The first argument is the finalizing continuation. The index $r$ specifies how many additional integer values are required to complete the evaluation to an integer; for example, $Ret_6\ n :: ExprRep_6\ [\,]$ will yield a result with no additional values needed, whereas $Sub_6 :: ExprRep_6\ [Integer, Integer]$ requires two additional integer values.

We can translate expressions into this representation:

$$
\begin{aligned}
&rep_6 : Expr \to ExprRep_6\ [\,]\\
&rep_6\ (Lit\ n) \quad\ \ = Ret_6\ n\\
&rep_6\ (Diff\ e\ e') = B_6^1\ (rep_6\ e)\ (B_6^2\ (rep_6\ e')\ Sub_6)
\end{aligned}
$$

which gives a tree as shown in figure 5(a). We can then evaluate the representative:

$$
\begin{aligned}
&abs_6 : ExprRep_6\ r \to (Integer \to Integer) \to Arrow\ r\ Integer\\
&abs_6\ (Ret_6\ n) = ret\ n\\
&abs_6\ Sub_6 \quad\ \ = sub\\
&abs_6\ (B_6^1\ x\ y) = b^1\ (abs_6\ x)\ (abs_6\ y)\\
&abs_6\ (B_6^2\ x\ y) = b^2\ (abs_6\ x)\ (abs_6\ y)
\end{aligned}
$$

yielding the interpretation shown in figure 5(b). The two steps constitute another evaluator:

$$
\begin{aligned}
&eval_6 : Expr \to Integer\\
&eval_6\ e = abs_6\ (rep_6\ e)\ halt
\end{aligned}
$$

which evidently deforests to $eval_5$. But this is still not promising as a compiler, because the intermediate representation $ExprRep_6$ is still tree-shaped, as figure 5 illustrates, whereas we were hoping for a linear sequence of instructions.

### 5.6 Linear code

Fortunately, associativity of generalized composition justifies rotation of the tree-shaped code in figure 5 into linear form, as shown in figure 6. Observe that

```
BRet7                            b0
 / \                             / \
3   BRet7                     ret   b1
     / \                      /     / \
    4   BSub7                3   ret   b0
          \                      /     / \
          BRet7                 4   sub   b1
           / \                            / \
          5   BSub7                    ret   b0
                \                      /     / \
                Halt7                 5   sub   halt
         (a)                          (b)
```

**Figure 6** (a) Linear code and (b) its interpretation

$$
\begin{aligned}
& eval_5\ e \\
= & \ [\![\ \text{definition}\ ]\!] \\
& eval'_5\ e\ halt \\
= & \ [\![\ b^0 \text{ is simply application}\ ]\!] \\
& b^0\ (eval'_5\ e)\ halt
\end{aligned}
$$

Then we can unroll the definition of $eval'_5$ on $expr$, as in figure 5(b), and exploit pseudo-associativity of the generalized compositions:

$$
\begin{aligned}
& eval_5\ expr \\
= & \ [\![\ \text{above}\ ]\!] \\
& b^0\ (eval'_5\ expr)\ halt \\
= & \ [\![\ \text{definition of } eval'_5\ ]\!] \\
& b^0\ (b^1\ (b^1\ (ret\ 3)\ (b^2\ (ret\ 4)\ sub))\ (b^2\ (ret\ 5)\ sub))\ halt \\
= & \ [\![\ \text{pseudo-associativity: } b^0\ (b^1\ h\ g)\ f = b^0\ h\ (b^0\ g\ f)\ ]\!] \\
& b^0\ (b^1\ (ret\ 3)\ (b^2\ (ret\ 4)\ sub))\ (b^0\ (b^2\ (ret\ 5)\ sub)\ halt) \\
= & \ [\![\ \text{pseudo-associativity: } b^0\ (b^1\ h\ g)\ f = b^0\ h\ (b^0\ g\ f)\ ]\!] \\
& b^0\ (ret\ 3)\ (b^0\ (b^2\ (ret\ 4)\ sub)\ (b^0\ (b^2\ (ret\ 5)\ sub)\ halt)) \\
= & \ [\![\ \text{pseudo-associativity: } b^0\ (b^2\ h\ g)\ f = b^1\ h\ (b^0\ g\ f)\ ]\!] \\
& b^0\ (ret\ 3)\ (b^1\ (ret\ 4)\ (b^0\ sub\ (b^0\ (b^2\ (ret\ 5)\ sub)\ halt))) \\
= & \ [\![\ \text{pseudo-associativity: } b^0\ (b^2\ h\ g)\ f = b^1\ h\ (b^0\ g\ f)\ ]\!] \\
& b^0\ (ret\ 3)\ (b^1\ (ret\ 4)\ (b^0\ sub\ (b^1\ (ret\ 5)\ (b^0\ sub\ halt))))
\end{aligned}
$$

The end result is shown in figure 6(b); note in particular that the generalized compositions now all nest to the right. In the same way as this specific example, pseudo-associativity allows us to rotate any tree-shaped evaluation into a linear form, in which no composition has another composition in its left argument. We can capture and enforce this property in another data representation:

$$
\begin{aligned}
& \textbf{data}\ ExprRep_7 : List\ Type \to Type\ \ \textbf{where} \\
& \quad Halt_7 : \qquad\qquad\qquad\qquad\qquad\qquad\quad ExprRep_7\ [Integer] \\
& \quad BRet_7 : Integer \to ExprRep_7\ (Integer :: r) \to ExprRep_7\ r \\
& \quad BSub_7 : ExprRep_7\ (Integer :: r) \to \qquad\quad\ ExprRep_7\ (Integer :: Integer :: r)
\end{aligned}
$$

Now a data structure of type $\mathit{ExprRep}_7\ r$ represents an evaluation function of type $\mathit{Arrow}\ r\ \mathit{Integer}$, and as before the index $r$ denotes the number of additional integer values required to complete the evaluation. But the datatype allows only a single ('right') child per non-leaf node. Note that in the process of rotating, the arities can get arbitrarily high—in particular, for right-nested expressions of the form $4-(3-(2-(1-1)))$ as in the 'subtractorial' function *subt* from section 2. The representation must therefore allow arbitrary indices, in contrast to $\mathit{ExprRep}_6$ which used indices only up to arity 2.

Given a function to append two representations:

$$
\begin{aligned}
&\mathit{append}_7 : \mathit{ExprRep}_7\ r \to \mathit{ExprRep}_7\ (\mathit{Integer} :: s) \to \mathit{ExprRep}_7\ (r +\!\!+ s)\\
&\mathit{append}_7\ \mathit{Halt}_7\ y \qquad = y\\
&\mathit{append}_7\ (\mathit{BRet}_7\ n\ k)\ y = \mathit{BRet}_7\ n\ (\mathit{append}_7\ k\ y)\\
&\mathit{append}_7\ (\mathit{BSub}_7\ k)\ y \ \ = \mathit{BSub}_7\ (\mathit{append}_7\ k\ y)
\end{aligned}
$$

we can transform from $\mathit{ExprRep}_6$ to $\mathit{ExprRep}_7$ by rotating the branching constructors $B_6^1, B_6^2$:

$$
\begin{aligned}
&\mathit{rotate}_7 : \mathit{ExprRep}_6\ r \to \mathit{ExprRep}_7\ r\\
&\mathit{rotate}_7\ (\mathit{Ret}_6\ n) = \mathit{BRet}_7\ n\ \mathit{Halt}_7\\
&\mathit{rotate}_7\ \mathit{Sub}_6 \qquad = \mathit{BSub}_7\ \mathit{Halt}_7\\
&\mathit{rotate}_7\ (B_6^1\ x\ y)\ = \mathit{append}_7\ (\mathit{rotate}_7\ x)\ (\mathit{rotate}_7\ y)\\
&\mathit{rotate}_7\ (B_6^2\ x\ y)\ = \mathit{append}_7\ (\mathit{rotate}_7\ x)\ (\mathit{rotate}_7\ y)
\end{aligned}
$$

But instead of going via the tree-shaped representation, we can convert expressions directly into the linear representation:

$$
\begin{aligned}
&\mathit{rep}_7 : \mathit{Expr} \to \mathit{ExprRep}_7\ [\,]\\
&\mathit{rep}_7\ (\mathit{Lit}\ n) \qquad = \mathit{BRet}_7\ n\ \mathit{Halt}_7\\
&\mathit{rep}_7\ (\mathit{Diff}\ e\ e') = \mathit{append}_7\ (\mathit{rep}_7\ e)\ (\mathit{append}_7\ (\mathit{rep}_7\ e')\ (\mathit{BSub}_7\ \mathit{Halt}_7))
\end{aligned}
$$

Either way, we can then interpret $\mathit{ExprRep}_7$ representations:

$$
\begin{aligned}
&\mathit{abs}_7 : \mathit{ExprRep}_7\ r \to \mathit{Arrow}\ r\ \mathit{Integer}\\
&\mathit{abs}_7\ \mathit{Halt}_7 \qquad\ \ = \mathit{halt}\\
&\mathit{abs}_7\ (\mathit{BRet}_7\ n\ k) = \mathit{ret}\ n\ (\mathit{abs}_7\ k)\\
&\mathit{abs}_7\ (\mathit{BSub}_7\ k)\ \ = \mathit{flip}\ (\mathit{sub}\ (\mathit{abs}_7\ k)) \qquad \text{-- note reversal of arguments again}
\end{aligned}
$$

Appending representations corresponds to composing their interpretations:

$$
b\ (\mathit{abs}_7\ y)\ (\mathit{abs}_7\ x) = \mathit{abs}_7\ (\mathit{append}_7\ x\ y)
$$

(note that backwards composition on the left-hand side becomes forwards sequencing on the right-hand side). This gives yet another implementation of the evaluator:

$$
\begin{aligned}
&\mathit{eval}_7 : \mathit{Expr} \to \mathit{Integer}\\
&\mathit{eval}_7\ e = \mathit{abs}_7\ (\mathit{rep}_7\ e)
\end{aligned}
$$

Since non-leaf nodes in the datatype $ExprRep_7$ have by construction a single child, this is a linear representation rather than a tree-shaped one, and there is an obvious compilation into the linear type of programs from before:

$$
\begin{array}{l}
compile_7 : Expr \to Prog_4 \\
compile_7 = compileRep_7 \cdot rep_7 \ \ \textbf{where} \\
\quad compileRep_7 : ExprRep_7\ r \to Prog_4 \\
\quad compileRep_7\ Halt_7 \qquad\quad = [\,] \\
\quad compileRep_7\ (BRet_7\ n\ k) = PushI\ n :: compileRep_7\ k \\
\quad compileRep_7\ (BSub_7\ k) \quad = SubI :: compileRep_7\ k
\end{array}
$$

Thus, $Halt_7$ corresponds to the empty program, and $BRet_7$ and $BSub_7$ each prefix one instruction onto a program.

The type $ExprRep_7\ r$ corresponds to programs which yield a single integer when run on an initial stack of shape $r$; there are no representations corresponding to incomplete programs such as $[PushI\ 3, PushI\ 4]$ which result in more than one value being pushed onto the stack.

## 6 Discussion

Reynolds [28] introduced the two-step process of conversion to continuation-passing style followed by defunctionalization of the continuations so introduced, in order to describe and classify interpreters used as a mechanism of language definition. Danvy and colleagues [13, 2] have extensively explored Reynolds's process, using it to explain the correspondence between various published interpreters and abstract machines, and to derive some new abstract machines from existing interpreters and vice versa. Our contribution has been to highlight the appeals to associativity in applications of this process, prominent in Wand's early application [31, 32] of Reynolds's approach but mostly unspoken since.

Reynolds and Danvy et al. focus on the lambda calculus as the defined language, using the process to explain the lambda calculus by implementing it in a simpler defining language. Subsequent work has focussed on the process, and has therefore chosen simpler defined languages than the lambda calculus as illustrations. In particular, a lot can be said in terms of a mere expression language consisting of numeric constants and addition, without touching the full Turing completeness of the lambda calculus. Hutton and colleagues have used this device in a series of papers [20, 5, 19, 6, 18]—so much so that McBride [24, 23] called it 'Hutton's Razor', although the device is actually much older than Hutton's uses [25, 32].

We've used Hutton's Razor ourselves in section 5, although we switched from addition to subtraction in order to be clear about where associativity is used. Following Reynolds's process leads to an abstract machine for evaluating arithmetic expressions, but not a compiler, because it cannot make a clear phase distinction between processing the input expression and performing arithmetic operations. Nevertheless, there is a well-known compilation scheme from expressions to a stack machine [25]. Bahr and Hutton [5] do manage to derive this compiler; but they do so by introducing stacks and stack transformers, motivated by operational reasoning:

> *The next step is to transform the evaluation function into a version that utilises a stack, in order to make the manipulation of argument values explicit. [5, §2.2]*

> *The next step is to transform [the evaluator] into CPS, in order to make the flow of control explicit. [5, §2.3]*

The operational justification ("in order to make … explicit") seems a pity, because CPS and defunctionalization already introduce stacks and make control flow explicit—these artifacts appear just by turning the handle, with no operational insight required. Our development in section 5 shows that the operational reasoning is not needed in order to arrive at a compiler. Instead, we got there by exploiting associativity of a generalized composition operator.

The observation that associativity of generalized composition leads directly to linearly structured target code is due to Wand [32], and explored in detail for more sophisticated languages by Henson [15, Chapter 8]. Wand also identifies the relevance of associativity for other continuation-based transformations [31].

Wand and Henson used an untyped setting; it seems that precise typing of the generalized composition requires (at least lightweight) dependent types, and can't otherwise be done statically in a type-safe manner, so there is no middle ground between these two positions. In fact, those dependent types express stack safety, making stack underflow a type error [26]. If one were to fully bite the dependently typed bullet, one could go further: to express not only the number and types of stack elements, but even their actual values [23, 27, 21], completing the proof of correctness of the compiler.

We have also shown that the same process applies to simpler problems than compilers, simpler even than compilers for basic expression languages. We showed in section 2 that it leads from the direct recursive implementation of factorial to the simple imperative loop for computing it—once associativity of multiplication has been exploited. Similarly, with associativity, it leads from the naive quadratic-time implementation of reverse to the linear-time accumulating-parameter version, essentially an application of Hughes's 'novel representation' [17] or difference lists in logic programming [11]. And again in conjunction with associativity, it leads to a linear-time tail-recursive tree traversal. Admittedly, there is a simpler linear-time tree traversal, if one dispenses with tail recursion [7, Exercise 7I]:

$$
\begin{array}{l}
\mathit{flatten}\ f = \mathit{flatten}'\ t\ [\,]\ \textbf{where} \\
\quad \mathit{flatten}'\ (\mathit{Tip}\ x)\quad \mathit{xs} = x :: \mathit{xs} \\
\quad \mathit{flatten}'\ (\mathit{Bin}\ t\ u)\ \mathit{xs} = \mathit{flatten}'\ t\ (\mathit{flatten}'\ u\ \mathit{xs})
\end{array}
$$

And if one follows the tree traversal development steps on the identity function on trees

$$\mathit{id} : \mathit{Tree}\ a \to \mathit{Tree}\ a$$

then the defunctionalized continuations *List* (*Either* (*Tree a*) (*Tree a*)) correspond to the *zipper* [16]. Do the same for the map function on trees

$treeMap : (a \to b) \to Tree\ a \to Tree\ b$

and the defunctionalized continuations *List* (*Either* (*Tree b*) (*Tree a*)) are tree *dissections* [22]—but neither of these exploits associativity.

The generalized composition we have used is in fact itself a special case of the yet more general notion of *operad*, "an abstraction of a family of composable functions of $n$ variables for various $n$, useful for the 'bookkeeping' and applications of such families" [30]. Our $b^r\ g\ f$ plugs an arity-$r$ function $f$ into an arity-1 function $g$, as shown in figure 3; more generally, there are $q$ different ways of plugging such an $f$ into an arity-$q$ function $g$, yielding an arity-$(q+r-1)$ function overall, and these again enjoy associativity in a natural way.

**Acknowledgements** I'm very grateful for constructive feedback from members of the Algebra of Programming research group at Oxford, especially Guillaume Boisseau, Josh Ko, Shin Cheng Mu, Richard Bird, Geraint Jones, Nick Wu, and Cezar Ionescu. I also owe thanks to members of IFIP Working Group 2.1 and participants at BOBKonf and Haskell Love for helpful questions and comments, to reviewers for constructive suggestions, to Mitch Wand for comments on his paper, and to Keith Clark and Tom Schrijvers for bibliographic assistance.

## A Appendix: A primer on Idris syntax

For the purposes of this paper, Idris is mostly used simply as a strongly typed pure functional programming language, similar in spirit to Haskell and Standard ML.

Syntactically, like both those languages, Idris uses square brackets $[1,2,3]$ for list values and $[\,]$ for the empty list. It uses Haskell's $+\!\!+$ rather than ML's @ for list append, Haskell's $\cdot$ rather than ML's $\circ$ for function composition, and Haskell's -- for line comments (which ML lacks). Lambda abstractions $\lambda x \Rightarrow e$ use Haskell's $\lambda$ (a backslash in ASCII) instead of ML's **fn**, but ML's $\Rightarrow$ instead of Haskell's $\rightarrow$. Case expressions **case** $x$ **of** $p \Rightarrow e \mid p' \Rightarrow e'$ also use ML's $\Rightarrow$ instead of Haskell's $\rightarrow$. Idris uses the ML convention of : for type declarations and :: for list cons, the opposite way round from the Haskell convention. Datatype declarations:

**data** *Tree a* = *Tip a* | *Bin* (*Tree a*) (*Tree a*)

use the same syntax as Haskell; but the list datatype is named (as in *List Integer*) rather than using brackets.

Each definition must have a type declaration, rather than allowing the type to be inferred as in ML and (most of the time) in Haskell. Functions must be defined before use; mutually recursive functions must be given in a **mutual** block:

$$
\begin{array}{l}
\mathit{parity} : \mathit{Nat} \rightarrow \mathit{Bool} \\
\mathit{parity}\ n = \mathit{odd}\ n\ \ \textbf{where} \\
\quad \textbf{mutual} \\
\qquad \mathit{odd}\ \ : \mathit{Nat} \rightarrow \mathit{Bool} \\
\qquad \mathit{odd}\ Z \qquad = \mathit{False} \\
\qquad \mathit{odd}\ (S\ n)\ \ = \mathit{even}\ n \\
\qquad \mathit{even} : \mathit{Nat} \rightarrow \mathit{Bool} \\
\qquad \mathit{even}\ Z \qquad = \mathit{True} \\
\qquad \mathit{even}\ (S\ n) = \mathit{odd}\ n
\end{array}
$$

Idris is dependently typed: types are first-class citizens, and the type of types is *Type*. So what would be a type synonym in Haskell:

$$\textbf{type}\ \mathit{FlatCont}_4\ a = [\mathit{Either}\ [a]\ (\mathit{Tree}\ a)]$$

is just an ordinary definition (of a function from types to types) in Idris:

$$
\begin{array}{l}
\mathit{FlatCont}_4 : \mathit{Type} \rightarrow \mathit{Type} \\
\mathit{FlatCont}_4\ a = \mathit{List}\ (\mathit{Either}\ (\mathit{List}\ a)\ (\mathit{Tree}\ a))
\end{array}
$$

We make only light use of the power of dependent types in this paper. In particular, here is the *Arrow* type from section 5:

$$
\begin{array}{l}
\mathit{Arrow} : \mathit{List}\ \mathit{Type} \rightarrow \mathit{Type} \rightarrow \mathit{Type} \\
\mathit{Arrow}\ [\,] \qquad b = b \\
\mathit{Arrow}\ (a :: as)\ b = a \rightarrow \mathit{Arrow}\ as\ b
\end{array}
$$

This is just another ordinary definition; it's a type-level function, but with dependent typing, types are values too. The function *Arrow* takes two arguments (a list of types and a single type), and returns a type as its result; the idea is that *Arrow as b* is the type of functions with argument types *as* and result type *b*. In Haskell it would require the lightweight dependently typed features of the GHC compiler—specifically, type-level analogues of value-level lists, and we would also need the corresponding type-level function to append type-level lists. In ML it could not be expressed at all. Generalized composition:

$$
\begin{array}{l}
b : \{as : \mathit{List}\ \mathit{Type}\} \rightarrow (b \rightarrow c) \rightarrow \mathit{Arrow}\ as\ b \rightarrow \mathit{Arrow}\ as\ c \\
b\ \{as = [\,]\} \qquad g\,f = g\,f \\
b\ \{as = _ :: _\}\ g\,f = b\ g \cdot f
\end{array}
$$

is defined by induction over the argument types index *as*; the curly brackets denote that *as* is an implicit argument—it could be inferred from the overall type (as $b, c$ are), but it made explicit to enable pattern matching on whether or not *as* is empty in the two clauses of the definition. In Haskell, it would require a family of singleton types in order to connect the value level with the type level; again, it is inexpressible in ML.

## About the author

**Jeremy Gibbons** is Professor of Computing at the University of Oxford, where he leads the *Algebra of Programming* research group.

**email:** jeremy.gibbons@cs.ox.ac.uk

**www:** http://www.cs.ox.ac.uk/jeremy.gibbons/

**orcid:** 0000-0002-8426-9917